\documentstyle[11pt,aaspp4]{article}
\input{psfig.sty}

\begin{document}

\title{Hard Burst Emission from the Soft Gamma Repeater SGR~1900+14}

\author{
Peter~M.~Woods\altaffilmark{1,3},
Chryssa~Kouveliotou\altaffilmark{2,3},
Jan~van~Paradijs\altaffilmark{1,4},
Michael~S.~Briggs\altaffilmark{1,3},
Kevin~Hurley\altaffilmark{5},
Ersin~{G\"o\u{g}\"u\c{s}}\altaffilmark{1,3},
Robert~D.~Preece\altaffilmark{1,3},
Timothy~W.~Giblin\altaffilmark{1,3},
Christopher~Thompson\altaffilmark{6}, and
Robert~C.~Duncan\altaffilmark{7}
}

\altaffiltext{1}{Department of Physics, University of Alabama in Huntsville, 
Huntsville, AL 35899; peter.woods@msfc.nasa.gov}
\altaffiltext{2}{Universities Space Research Association}
\altaffiltext{3}{NASA Marshall Space Flight Center, SD50, Huntsville, AL
35812}
\altaffiltext{4}{Astronomical Institute ``Anton Pannekoek'', University of 
Amsterdam, 403 Kruislaan, 1098 SJ Amsterdam, NL}
\altaffiltext{5}{University of California at Berkeley, Space Sciences
Laboratory, Berkeley, CA 94720-7450}
\altaffiltext{6}{Department of Physics and Astronomy, University of North
Carolina, Philips Hall, Chapel Hill, NC 27599-3255}
\altaffiltext{7}{Department of Astronomy, University of Texas, RLM 15.308,
Austin, TX 78712-1083}

\begin{abstract}

We present evidence for burst emission from SGR~1900$+$14 with a power-law high
energy spectrum extending beyond 500 keV.  Unlike previous detections of high
energy photons during bursts from SGRs, these emissions are not associated with
high-luminosity burst intervals.  Not only is the emission hard, but the
spectra are better fit by Band's GRB function rather than by the traditional
optically-thin thermal bremsstrahlung model.  We find that the spectral
evolution within these hard events obeys a hardness/intensity {\it
anti}-correlation.  Temporally, these events are distinct from typical SGR
burst emissions in that they are longer ($\sim$ 1 s) and have relatively smooth
profiles.  Despite a difference in peak luminosity of $\gtrsim 10^{11}$ between
these bursts from SGR~1900$+$14 and cosmological GRBs, there are striking
temporal and spectral similarities between the two kinds of bursts, aside from
spectral evolution.  We outline an interpretation of these events in the
context of the magnetar model.

\end{abstract}


\section{Introduction}

Soft gamma repeaters (SGRs) constitute a group of high-energy transients named
for the observed characteristics which set them apart from classical Gamma-Ray
Bursts (GRBs).  SGRs emit brief ($\sim$ 0.1 sec), intense (up to 10$^{3}$ --
10$^{4}$ L$_{\rm Edd}$) bursts of low-energy $\gamma$-rays with recurrence
times which range from seconds to years (Kouveliotou 1995).  The vast majority
of SGR burst spectra ($\gtrsim$ 20 keV) can be fit by an Optically-Thin Thermal
Bremsstrahlung (OTTB) model with temperatures between 20 and 35 keV (Fenimore,
Laros, \& Ulmer 1994; {G\"o\u{g}\"u\c{s}} et al. 1999).  These spectra show
little or no variation over a wide range of time scales (Fenimore et al. 1994),
both within individual bursts, and between source active periods which cover
years.  There have been some exceptions, however, where modest hard-to-soft
evolution within bursts from SGR~1806$-$20 was detected (Strohmayer \& Ibrahim
1997).

During the past 20 years, two giant flares have been recorded from two of the
four known SGRs: one from SGR~$0526-66$ on 5 March 1979 (Mazets et al. 1979),
and one from SGR~1900$+$14 on 27 August 1998 (Hurley et al. 1999a).  These
flares differ from the more common bursts in many ways.  They are far more
energetic (by 3 orders of magnitude in peak luminosity), persist for hundreds
of seconds during which their emission is modulated at a period that reflects
the spin of an underlying neutron star, and have harder initial spectra.  The
hard spectra of the peak of these flares have OTTB temperatures 200 -- 500 keV
(Fenimore et al. 1991; Hurley et al. 1999a; Mazets et al. 1999a), although this
model should not necessarily be taken as an accurate valid description of these
spectra as severe dead-time problems for most instruments limited the efficacy
of spectral deconvolution.  Feroci et al. (1999) require the first $\sim$ 67 s
of the August 27$^{\rm th}$ flare be fit with a two-component spectrum,
consisting of an OTTB (31 keV) and a power-law (-- 1.47 photon index).  Hard
burst emission has also been detected during the brightest burst recorded from
the newly discovered SGR~1627$-$41 (Woods et al. 1999a; Mazets et al. 1999b). 
Further evidence for hard emission from SGRs comes from RXTE observations of
SGR~1806$-$20.  For a small fraction of the more common short events, high OTTB
spectral temperatures in the range 50 -- 170 keV were measured (Strohmayer \&
Ibrahim 1997).  

Here, we present strong evidence for spectrally hard burst emission from
SGR~1900$+$14 during its recent active episode which started in May 1998.  Two
events recorded with BATSE which are positionally consistent with SGR~1900$+$14
and temporally coincident with the recent active period of the source, show
temporal and spectral signatures quite distinct from typical SGR burst
emissions.  We show that although the time-integrated spectrum of each event
resembles a classical GRB spectrum, spectral evolution is found which obeys a
hardness/intensity anti-correlation, never before seen in GRBs.

\section{Burst Association with SGR~1900+14}

On 22 October 1998, during a period of intense burst activity of SGR~1900$+$14
(Woods et al. 1999b), BATSE triggered at 15:40:47.4 UT on a $\sim$ 1 s burst
with a smooth, FRED-like (Fast Rise Exponential Decay) temporal profile which
is commonly seen in GRB light curves, but is rare for SGR events.  This burst
was located near SGR~1900$+$14 (Figure 1), but was longer than typical bursts
from this source and its spectrum was much harder\footnote{The similarities of 
this event with GRBs in spectrum and temporal structure was noted independently
by D. Fargion (1999)} (Figure 2a).  Using Ulysses and BATSE, an IPN annulus was
constructed and the joint BATSE/IPN error box (Figure 1) contained the known
source location of SGR~1900$+$14 (Frail, Kulkarni \& Bloom 1999).  

\begin{figure}[htb]
\centerline{
\psfig{file=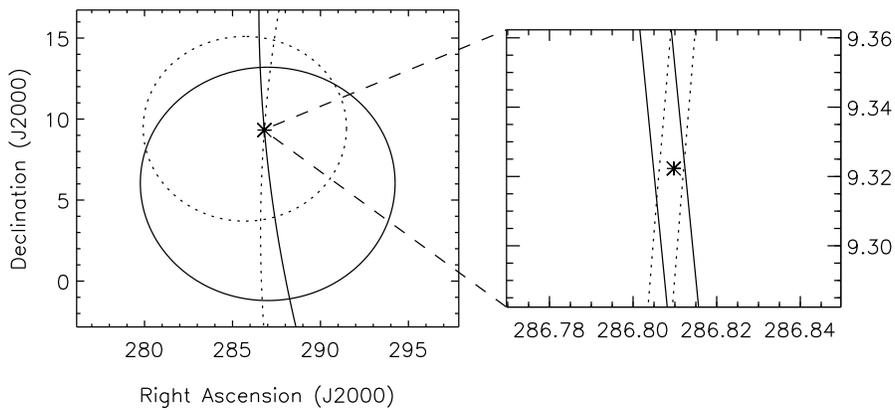,width=5.0in}}
\vspace{-0.7in}
\caption{Localizations of 981022 (dotted) and 990110 (solid) with BATSE
(circles denote 90\% statistical + systematic error radii) and BATSE/Ulysses
IPN arcs (99\% statistical + systematic error).  VLA location of SGR~1900$+$14
is denoted by the asterisk. \label{fig:loc_hb}}
\vspace{11pt}
\end{figure}

Without invoking any assumptions about source activity, we estimate the
probability that a GRB within the BATSE database with an IPN arc would contain
any known SGR by chance.  For our purposes, we will assume GRBs are isotropic
and the angular size of the known SGR error boxes are small compared to the
joint BATSE/IPN error box.  The probability $p$ reduces to $p
\approx~(1/4\pi)~NA$ where $N$ is the number of known SGRs and $A$ is the area
of the burst error box in steradians.  With four known SGRs, we find a chance
probability of 3 $\times$ 10$^{-6}$ that a GRB with the given error box area
will overlap a known SGR.  We now normalize this probability by multiplying by
the number of trials (i.e. the number of BATSE/Ulysses IPN arcs as of March
1999, which is 641), which gives the probability of a chance association of 2
$\times$ 10$^{-3}$.  The burst was detected during a period when SGR~1900$+$14
was burst active (a fact not used in the probability calculation), further
strengthening the association.

\begin{figure}[htb]
\centerline{
\psfig{file=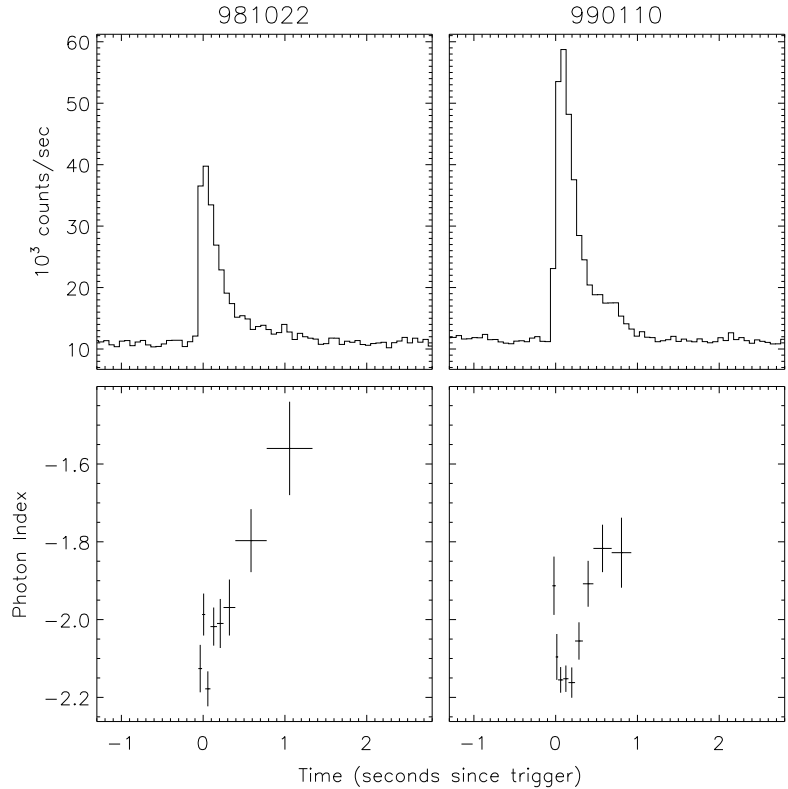,height=4.5in}}
\vspace{-0.15in}
\caption{light curves (25 $-$ 2000 keV) of two bursts from SGR~1900$+$14
detected with BATSE.  BATSE trigger 7171 (981022) is shown in panel (a) and
trigger 7315 in panel (b).  Panels (c) and (d) give the photon index as a
function of time to illustrate the spectral evolution observed during these
events. \label{fig:lc_hb}}
\vspace{11pt}
\end{figure}

Ten weeks later on 10 January 1999 at 08:39:01.4 UT, a strikingly similar burst
(Figure 2b) was recorded with BATSE, which again located near SGR~1900$+$14. 
This burst also triggered Ulysses, so an IPN annulus was constructed which
again contained the position of SGR~1900$+$14 (Figure 1).  Using the same
arguments as before, we find an upper limit to the  probability that the burst
and any known SGR are related by chance coincidence of 3 $\times$ 10$^{-3}$. 
The combined probability (product of the two individual probabilities) that
these two events are GRBs with BATSE/IPN error boxes that are consistent with a
known SGR by chance coincidence is 6 $\times$ 10$^{-6}$.

An alternative possibility is that these two bursts constitute two
gravitationally lensed images of the same GRB (Paczy\'nski 1986).  However, we
consider this highly unlikely given the positional coincidence of
SGR~1900$+$14, the temporal correlation with a known burst active period for
the source, and the anti-correlation between hardness and intensity (see
section 3).  We conclude that these two bursts originated from SGR~1900$+$14.

\section{Burst Spectra}

To fit the time-integrated spectrum for each burst, we used High Energy
Resolution Burst (HERB) data which have 128 energy channels covering 20 -- 2000
keV (see Fishman et al. 1989 for a description of BATSE data types).  We fit a
third-order polynomial to approximately 300 sec of pre-burst and post-burst
data and interpolated between these intervals to estimate the background at the
time of the burst.  This background was subtracted and we fit the resulting
burst spectrum using WINGSPAN (WINdow Gamma SPectral ANalysis) to three models:
an OTTB (dN/dE $\propto$ E$^{-1}$ exp[-E/$kT$]), a simple power-law, and Band's
GRB function (Band et al. 1993).  We find that for each burst, the spectrum is
not well characterized by the OTTB model based upon the large value of
${\chi}^{2}_{\nu}$ (Table 1).  Using a $\Delta\chi^2$ test between the Band and
OTTB models, we find the Band function is strongly favored over the OTTB, with
probabilities of 4.6~$\times$~10$^{-7}$ and 7.2~$\times$~10$^{-11}$ that these
$\chi^2$ differences would occur by chance for the respective bursts. 
Furthermore, the Band function is favored over the simple power-law for the
second, brighter event with a significance level of 1.3~$\times$~10$^{-3}$,
although inclusion of a low-energy cutoff with the power-law model eliminates
this advantage.  Figure 3 shows the data and folded Band model for each
burst.  

\begin{figure}[htb]
\centerline{
\psfig{file=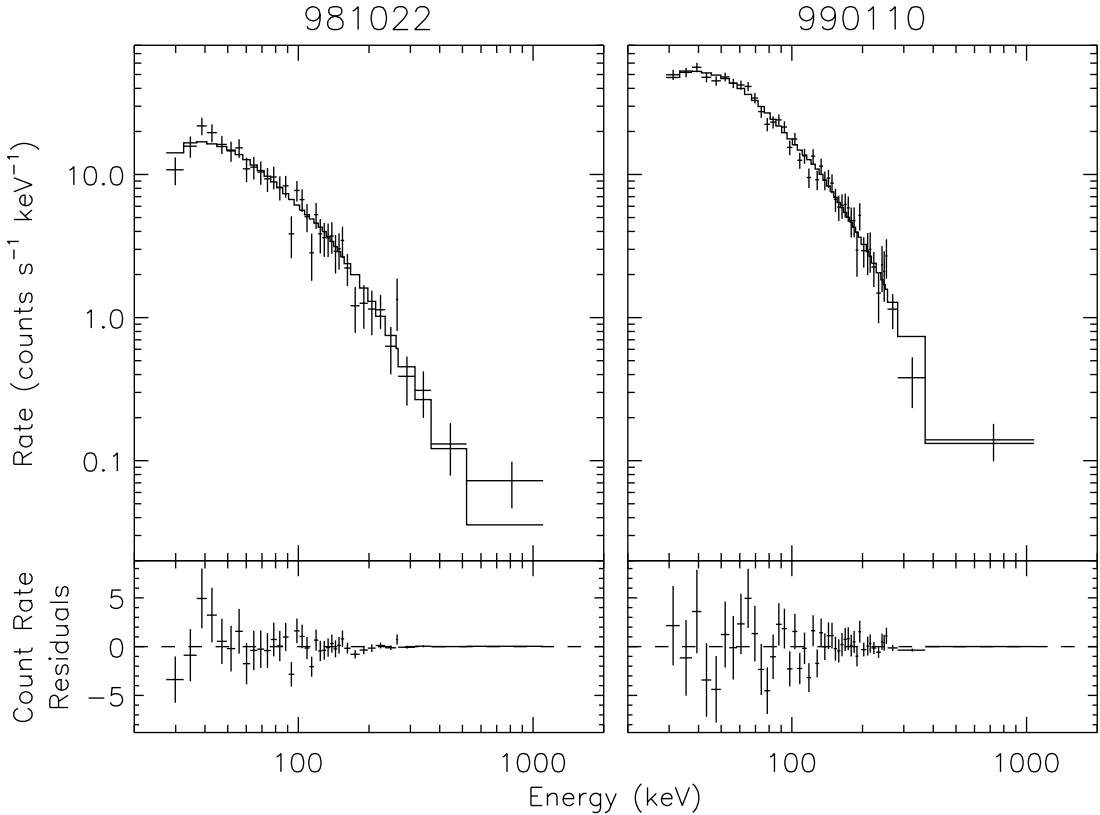,height=4.5in}}
\vspace{-0.1in}
\caption{Count spectrum of 981022 (a) and 990110 (b) as fit to Band's GRB
function.  See Table 1 for spectral fit information.  The spectrum has been
binned for display purposes. \label{fig:spec_hb}}
\vspace{11pt}
\end{figure}

For comparison to past results, we have included the OTTB fit parameters (Table
1) to clearly demonstrate the difference between the spectra of these bursts
and typical SGR burst emissions.  Specifically, for SGR~1900$+$14 during its
recent active episode, a weighted mean of 25.7 $\pm$ 0.8 keV was found for the
OTTB temperature of 22 events detected with BATSE ({G\"o\u{g}\"u\c{s}} et al.
1999).  Clearly, the temperatures found for these two bursts are much higher
than these typical values.

\placetable{tbl-1}

\begin{center}
\begin{deluxetable}{ccccc}
\small
\tablecaption{Spectral fit summary \label{tbl-1}}

\tablehead{
\colhead{Burst}  &  
\colhead{Model}    &  
\colhead{$\chi^2$/dof}  &  
\colhead{$kT$ or E$_{\rm peak}$ (keV)}     &  
\colhead{Photon index\tablenotemark{a}}
}

\startdata

981022  &  OTTB       &  119.4/96  &  102 $\pm$ 5    &  $-$                 \nl

        &  Power law  &   92.7/96  &  $-$            &  -- 1.91 $\pm$ 0.06  \nl

        &  Band's GRB &   90.2/94  &  54 $\pm$ 50    &  -- 1.96 $\pm$ 0.08  \nl

990110  &  OTTB       &  139.8/95  &  94 $\pm$ 4     &  $-$                 \nl

        &  Power law  &  106.4/95  &  $-$            &  -- 2.06 $\pm$ 0.03  \nl

        &  Band's GRB &   93.1/93  &  59 $\pm$ 11    &  -- 2.19 $\pm$ 0.06  \nl

\enddata

\tablenotetext{a}{For Band's GRB function, the high energy index $\beta$ is
given here.  The low energy index $\alpha$ is not well determined due to the
low E$_{\rm peak}$ values.}

\end{deluxetable}
\end{center}

Using data with coarser spectral resolution, but finer time resolution, we fit
multiple segments of each burst in order to search for spectral evolution.  For
the initial rise of each event, we used Time-Tagged Event (TTE) data which have
only 4 energy channels but 2 $\mu$s time resolution.  For the peak and tail, we
used Medium-Energy Resolution (MER) data which have 16 energy channels and 16
ms time resolution.  In order to sustain good signal-to-noise for a reasonable
parameter determination, the bursts were divided into 8 and 9 intervals,
respectively.  Guided by our fit results of the time-integrated spectra and our
limited number of energy channels, we chose to fit the power-law model to these
spectra.  We find significant spectral evolution through each burst as the
power-law photon index varies between -- 1.5 and -- 2.4 (Figures 2c and 2d). 
We note a general soft-to-hard trend in the time evolution of the spectra of
these bursts.  This appears to be a consequence of a relatively faster temporal
rise than decay and an intensity/hardness anti-correlation for these events
(Figure 4).  To quantify the significance of this correlation, we calculated
the Spearman rank-order correlation coefficient ($\rho = - 0.86$) between the
energy flux and photon index.  The probability of obtaining a coefficient of
this value from a random data set is 8.3 $\times$ 10$^{-6}$, thus the
anti-correlation is significant.

\begin{figure}[htb]
\centerline{
\psfig{file=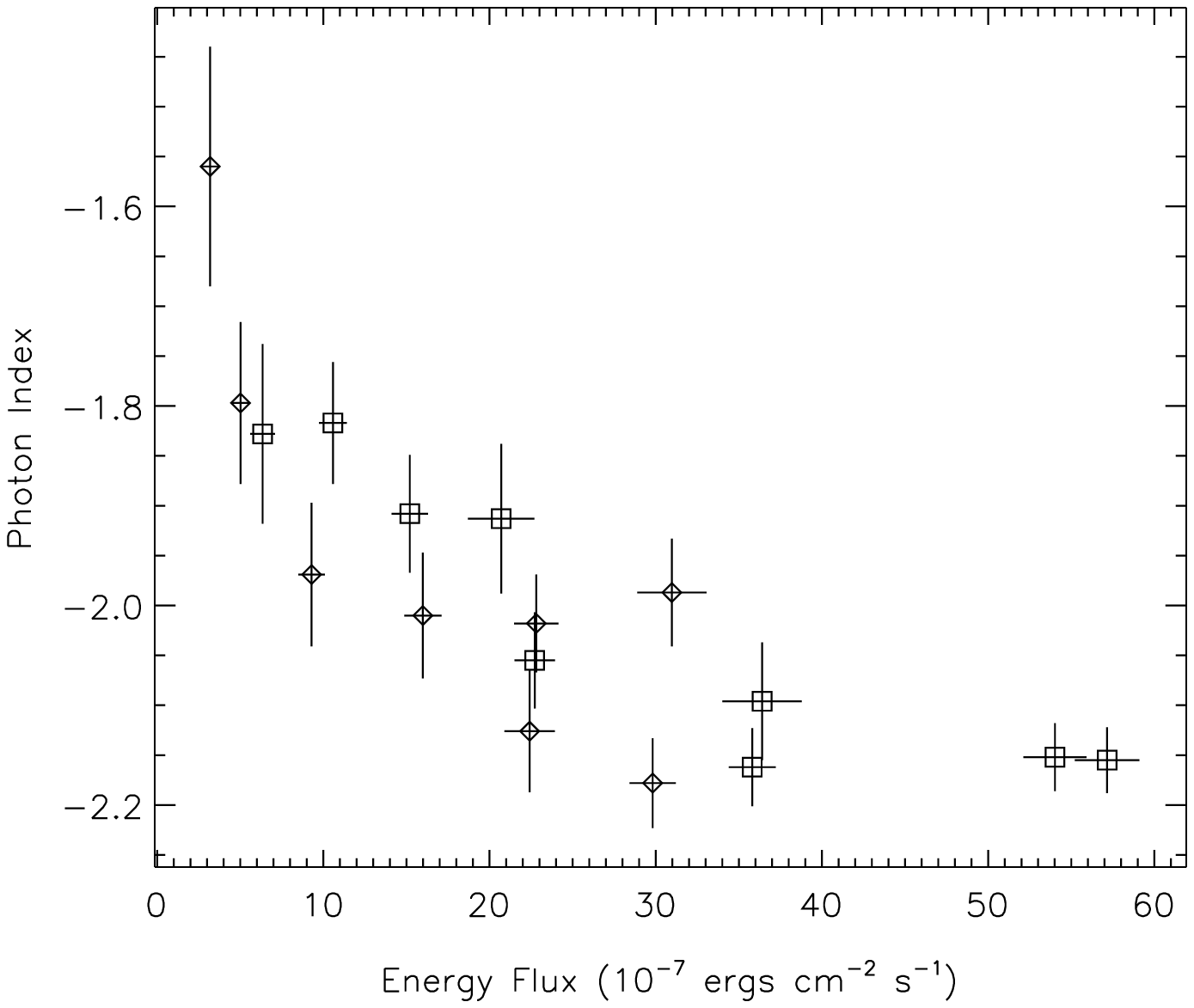,height=3.5in}}
\vspace{0.0in}
\caption{Energy flux vs. photon index for 981022 (diamonds) and 990110
(squares). \label{fig:spec_hb}}
\vspace{11pt}
\end{figure}

The peak fluxes and fluences of the two events are not exceptional when
compared to other burst emissions from this source.  We find peak fluxes (0.064
s timescale) of (2.94~$\pm$~0.15) and (4.96~$\pm$~0.18) $\times$ 10$^{-6}$ ergs
cm$^{-2}$ s$^{-1}$ and fluences (25 -- 2000 keV) of (1.14~$\pm$~0.04) and
(1.85~$\pm$~0.04) $\times$ 10$^{-6}$ ergs cm$^{-2}$ for the bursts of 981022
and 990110, respectively.  For a distance of 7 kpc (Vasisht et al. 1994), these
correspond to peak luminosities 1.7 and 2.9 $\times$ 10$^{40}$ ergs s$^{-1}$
and burst energies 0.65 and 1.1 $\times$ 10$^{40}$ ergs (assuming isotropic
emission).  The ranges of peak fluxes and fluences for SGR~1900$+$14 bursts
observed recently with BATSE are $0.3~-~20$ $\times$ 10$^{-6}$ ergs cm$^{-2}$
s$^{-1}$, and $0.02~-~25$ $\times$ 10$^{-6}$ ergs cm$^{-2}$, respectively
({G\"o\u{g}\"u\c{s}} et al. 1999).  The measured values for these bursts with
harder spectra are well within the corresponding observed ranges.

\section{A Comparison with Observed GRB Characteristics}

For each event, we have calculated two quantities that are traditionally used
to delineate between the two classes (Kouveliotou et al. 1993) of GRBs in the
BATSE catalog, specifically t$_{90}$ and spectral hardness (Ch 3/Ch 2).  We
find t$_{90}$ durations of 1.2 $\pm$ 0.2 and 0.9 $\pm$ 0.2 s, and fluence
hardness ratios (3/2) of 1.9 $\pm$ 0.1 and 2.1 $\pm$ 0.1 for the bursts of
981022 and 990110, respectively.  When plotted together with the reported
values of the 4Br catalog (Paciesas et al. 1999), we find these two bursts fall
outside the main concentrations of each distribution (i.e. the long, soft class
and the short, hard class), but nearer the centroid of the short, hard class. 
We are currently looking more closely at events in the same region of this
diagram that were classified as GRBs.

Given the spectral similarities between these two events and a fair fraction of
GRBs, the time-integrated spectrum is not sufficient to distinguish these
bursts from GRBs.  Many GRBs show a strong hardness/intensity correlation
within individual bursts (Ford et al. 1995; Preece et al. 1999), but to the
best of our knowledge, a consistent {\it anti}-correlation throughout a GRB has
never been seen.  The two bursts from SGR~1900$+$14 do, however, show a strong
hardness/intensity anti-correlation (Figure 4).  If this behavior is inherent
to all {\it hard} SGR events, it would be a useful diagnostic (but secondary to
location) with which to select them.

\section{Discussion}

We have shown strong evidence for hard emission from SGR~1900$+$14 during two
bursts of average intensity as observed with BATSE.  It is clear that this type
of burst emission in SGR~1900$+$14 is rare (1\% of the SGR~1900$+$14 events
acquired with BATSE during the 1998-1999 active period).  Their occurrence
following the August 27 flare may suggest a causal relationship, but this is
difficult to pin down given the rarity of hard events and the global enhanced
burst activity at the time. The clear distinction between these two events in
hardness, spectral form, spectral evolution, duration, and lack of temporal
variability suggests these bursts are created either by physical processes
different from those which produce the more common, soft events, or in a region
whose ambient properties that effect the emitted radiation (e.g. magnetic
field, optical depth, etc.) are different. 

The proposed identification of SGRs with very strongly magnetized neutron stars
(Duncan \& Thompson 1992) has received strong support from the discovery that
both SGR~1806$-$20 and SGR~1900$+$14 are X-ray pulsars spinning down at rapid
rates (Kouveliotou et al. 1998, 1999;  Hurley et al. 1999b).  In this magnetar
model, the typically soft SGR bursts are explained in the following way.   The
internal magnetic field is strong enough to diffuse rapidly through the core,
thereby stressing the crust (Thompson \& Duncan 1996).  A sudden fracture
injects a pulse of Alfv\'en radiation directly into the magnetosphere, which
cascades to high wavenumber and creates a trapped fireball (Thompson \& Duncan
1995;  Thompson \& Blaes 1998).  The soft spectrum arises from a combination of
photon splitting and Compton scattering in the cool, matter-loaded envelope of
the fireball.

The relative hardness of the two reported events, combined with luminosities
in excess of $10^{40}$ erg s$^{-1}$, points directly to an emission region of
shallower scattering depth, $\tau_{\rm es} < 1$,  situated outside $\sim
10^3(L_X/10^{40}~{\rm erg~s^{-1}})$ neutron star radii. Inverse Compton
emission is suggested by the lack of a correspondence between the spectral
break energy and a cooling energy at this large a radius. For example, bulk
Alfv\'en motions are an effective source of Compton heating even in the absence
of Coulomb coupling between electrons and ions (Thompson 1994).  In certain
circumstances, one expects  that Alfv\'en radiation will disperse rapidly
throughout the magnetosphere: if the initial impulse occurs on extended dipole
field lines; or if it involves a buried fracture of the crust.  This rapid
dispersal is made possible by the strong coupling between external Alfv\'en
modes and internal seismic waves (cf. Blaes et al. 1989).  The wave energy is
then distributed  logarithmically with radius, with the wave amplitude $\delta
B/B$ approaching unity at the Alfv\'en radius $R_A/R_\star \simeq
900\,(B_\star/4.4\times 10^{14}~{\rm G})^{1/2}\, (L_A/10^{40}~{\rm
erg~s^{-1}})^{-1/4}$ (Thompson \& Blaes 1998).  Here $B_\star$ is the polar
dipole magnetic field, $L_A$ is the luminosity in escaping waves, and $R_\star
\simeq 10$ km.  Damping by leakage onto open field lines (which extend beyond
the Alfv\'en radius) occurs on a timescale $t_{\rm damp} =
3(2R_A/R_\star)(R_\star/c) = 0.2\,(B_\star/4.4\times 10^{14}~{\rm G})^{1/2}\,
(L_A/10^{40}~{\rm erg~s^{-1}})^{-1/4}$ s. This lies close to the FWHM of the
two reported bursts.  Wave excitations near the neutron star undergo a
turbulent cascade on a similar timescale, and will generate softer seed X-ray
photons (Thompson et al. 1999).

Independent of the detailed physical mechanism, these observations demonstrate
that a Galactic source -- probably a strongly magnetized neutron star with a
large velocity -- is capable of producing a burst of $\gamma$-rays whose
time-integrated spectrum resembles that of a classical GRB. The similarity is
remarkable in light of the difference in peak luminosities of $\gtrsim 10^{11}$
(modulo beaming factors).  However, the two bursts from SGR~1900$+$14 presented
here are by no means `typical' GRBs, given their low E$_{\rm peak}$ values,
unusual spectral evolution, and their position in the duration-hardness plane.  
Furthermore, we know the currently active SGRs in our Galaxy lie close to the
Galactic plane ($z_{\rm rms}$ $\simeq$ 66~pc), so their contribution to the
BATSE GRB catalog must be minimal on account of the isotropic spatial
distribution of GRBs.  A larger contribution from older magnetars that have
moved away from the Galactic plane is conceivable, but it would require that
the preponderance of hard bursts to soft bursts increases tremendously with
age.

\acknowledgments{\noindent {\it Acknowledgements} -- We thank Chip Meegan for
useful discussions.  PMW acknowledges support under the cooperative agreement
NCC 8-65.  JvP acknowledges support under NASA grants NAG 5-3674 and NAG
5-7060.  KH is grateful for support under JPL Contract 958056 and NASA grant
NAG5-7810.  CT acknowledges support from the Alfred P. Sloan Foundation.}

\end{document}